\documentclass[onecolumn,showpacs,superscriptaddress,10pt]{revtex4-1} 
\usepackage{amsmath,amssymb,graphicx,subfigure}

\usepackage{color}

\begin{document}
\title{Dual Hg-Rb magneto-optical trap}
\author{Marcin Witkowski}\email{Corresponding author: mwitkowski@uni.opole.pl}
\affiliation{Institute of Physics, Faculty of Physics, Astronomy and Informatics, Nicolaus Copernicus University, Grudzi\c{a}dzka 5, PL-87-100 Toru\'n, Poland}
\affiliation{Institute of Physics, University of Opole, Oleska 48, PL-45-052 Opole, Poland}
\author{Bart\l{}omiej Nag\'orny}
\affiliation{Institute of Physics, Faculty of Physics, Astronomy and Informatics, Nicolaus Copernicus University, Grudzi\c{a}dzka 5, PL-87-100 Toru\'n, Poland}
\affiliation{Time and Frequency Department, Astrogeodynamic Observatory of Space Research Center, Borowiec, Drapa\l{}ka 4, PL-62-035 K\'ornik,  Poland}
\author{Rodolfo Munoz-Rodriguez}
\affiliation{Institute of Physics, Faculty of Physics, Astronomy and Informatics, Nicolaus Copernicus University, Grudzi\c{a}dzka 5, PL-87-100 Toru\'n, Poland}
\author{Roman Ciury\l{}o}
\affiliation{Institute of Physics, Faculty of Physics, Astronomy and Informatics, Nicolaus Copernicus University, Grudzi\c{a}dzka 5, PL-87-100 Toru\'n, Poland}
\author{Piotr Szymon \.Zuchowski}
\affiliation{Institute of Physics, Faculty of Physics, Astronomy and Informatics, Nicolaus Copernicus University, Grudzi\c{a}dzka 5, PL-87-100 Toru\'n, Poland}
\author{S\l{}awomir Bilicki}
\affiliation{Institute of Physics, Faculty of Physics, Astronomy and Informatics, Nicolaus Copernicus University, Grudzi\c{a}dzka 5, PL-87-100 Toru\'n, Poland}
\affiliation{LNE-SYRTE, Observatoire de Paris, PSL Research University, CNRS, Sorbonne Universit\'es, UPMC Univ. Paris 06, 61 avenue de l'Observatoire, 75014 Paris, France}
\author{Marcin Piotrowski}
\affiliation{Institute of Physics, Faculty of Physics, Astronomy and Informatics, Nicolaus Copernicus University, Grudzi\c{a}dzka 5, PL-87-100 Toru\'n, Poland}
\affiliation{Commonwealth Scientific and Industrial Research Organisation, Manufacturing Flagship, Pullenvale QLD 4069 Australia}
\author{Piotr Morzy\'nski}
\affiliation{Institute of Physics, Faculty of Physics, Astronomy and Informatics, Nicolaus Copernicus University, Grudzi\c{a}dzka 5, PL-87-100 Toru\'n, Poland}
\author{Micha\l{} Zawada}
\affiliation{Institute of Physics, Faculty of Physics, Astronomy and Informatics, Nicolaus Copernicus University, Grudzi\c{a}dzka 5, PL-87-100 Toru\'n, Poland}

%
%

\begin{abstract}
  We present a two-species laser cooling apparatus capable of simultaneously collecting Rb and Hg atomic gases into a magneto-optical trap (MOT). The atomic sources, laser system, and vacuum set-up are described. While there is a loss of Rb atoms in the MOT due to photoionization by the Hg cooling laser, we show that it does not prevent simultaneous trapping of Rb and Hg. We also demonstrate interspecies collision-induced losses in the $^{87}$Rb-$^{202}$Hg system.
\end{abstract}


\maketitle 


\section{Introduction}

Simultaneous laser cooling of multiple species has attracted a great deal of attention in recent years. Such mixtures are a starting point toward production of cold polar molecules~\cite{Quemener2012} having numerous potential applications in quantum information theory~\cite{DeMille2002,Buchler2007,Baranov2012}, high precision spectroscopy~\cite{Zelevinsky2008,Hudson2006} and fundamental studies of new physics.
In 2008 it was demonstrated that ground state ultracold KRb molecules can be produced by magnetoassociation followed by STIRAP (stimulated Raman adiabatic passage) ~\cite{Ni231} and since then several other molecules such as RbCs, NaK, NaRb have been produced employing similar techniques~\cite{ PhysRevLett.113.205301,PhysRevLett.113.255301,PhysRevLett.114.205302,PhysRevLett.116.205303}. Ultracold samples of molecules can also be created as a product of spontaneous decay of  photoassociated ultracold atoms, with the notable example of LiCs molecules~\cite{Deiglmayr2008}.

A very interesting class of molecules can be produced from alkali and closed-shell alkaline-earth(like) atoms~\cite{Zuchowski2010}. These open-shell molecules have both electric- and magnetic-dipole moments which opens up new possibilities of their control by external fields. In particular Micheli et al.~\cite{Micheli2006} proposed new quantum simulators based on such molecules.
There were also proposals for applications of such molecules in high-sensitivity imaging of magnetic fields~\cite{Alyabyshev2012} as well as high-precision measurements of electron-to-proton mass ratio~\cite{Kajita2009,Kajita2011} and applications in searches of limits of electric dipole moment of electron~\cite{Meyer2006}.
Moreover, alkaline-earth(like) atoms are of great importance in ultracold physics and optical metrology. They are crucial constituents of optical atomic clocks~\cite{Ludlow2015} which are the most accurate and precise scientific instruments.

Since the first demonstration of two-species MOT of $^{85}$Rb and $^{87}$Rb \cite{Suptitz:94}, various different mixtures have been studied experimentally in MOTs, including alkali metal atoms, metastable-states systems~\cite{Sukenik2002}, highly magnetic mixtures~\cite{Hensler2004} or three-species MOTs~\cite{Taglieber2006}.
However, there are only few  systems to date with mixtures of alkali and alkaline-earth(like) atoms. The group of G\"{o}rlitz reported several studies of simultaneous   trapping of Rb and Yb atoms in combined magnetic and optical traps \cite{Baumer2011, Tassy2010}, and of photoassociation spectroscopy~\cite{Borkowski2013,Munchow2011,Nemitz2009}. More recently, a quantum degenerate mixture of fermionic Yb with Rb has been studied \cite{Vaidya2015}. Another important system in which co-trapping atoms with different electronic spins was reported is the mixture of ultracold Sr and Rb atoms \cite{Aoki2013,Pasquiou2013}. In particular, Pasquiou et al.  achieved quantum degeneracy in both species. For the YbLi system two groups achieved quantum degenerate state ~\cite{Hansen2013,Hara2011} and  studies of photoassociation were performed \cite{Roy2016}. Finally, it is worth mentioning that Kemp et al. have recently reported a double MOT setup for co-trapping the Yb and Cs atoms~\cite{Kemp2016}.

In this article, we report the first experimental system for simultaneous cooling and trapping of Hg and Rb atoms in the MOT. Rubidium has its cooling transition (780~nm) at a wavelength convenient for widely available high power diode lasers. It also has a moderate temperature required to obtain vapour pressures at the level required in MOTs. These features make Rb one of the most commonly used atomic species employed for laser cooling. On the other hand, the cooling transition for mercury atoms (253.7~nm) belongs to the UV-C range which is not covered by commercially available diode laser sources.
Nevertheless, there are several features that make mercury atoms particularly interesting. Mercury has seven stable isotopes, two fermionic and five bosonic. 
It is regarded as the best candidate for the most accurate optical atomic clock~\cite{Yi2011,PhysRevLett.100.053001}. This is due to its low sensitivity to black body radiation shift which is one of the fundamental limitations of the performance of the atomic clocks~\cite{Safronova2013}.
At this point, however, because of the relatively less complicated experimental set-up, i.a. thanks to required laser wavelengths, most accurate optical lattice clocks are based on strontium and ytterbium. Nevertheless, to achieve best accuracy in Sr and Yb optical clocks a cryogenic vacuum chamber is usually required. 
The frequency uncertainty of the ${}^{199}$Hg optical lattice clock  is presently limited by the lattice light shift \cite{Yamanaka2015}.
In such a clock, another advantage of mercury, i.e. high sensitivity to fine-structure constant allows finding new limits on its variation in time~\cite{Flambaum2009,Angstmann2004}, or on coupling between Dark Matter and Standard Model~\cite{Essig2013,Derevianko2014,Wcislo2016}.
Moreover, due to its very high nuclear charge, mercury is an appropriate candidate to perform fundamental experiments on parity-violation~\cite{Bouchiat1977} or look for physics beyond the Standard Model, including CP-violating permanent electric dipole moment and CP-violating polarizabilities \cite{PhysRevA.72.012101}. In particular, Meyer and Bohn suggested that molecules containing mercury were good candidates for experimental searches for electric dipole moment of the electron~\cite{Meyer2009}.

Finally, let us point out that the van der Waals interaction of Hg with other atoms is significantly weaker in comparison to Sr or Yb. Together with other advantages mentioned above it makes molecules containing Hg especially suitable for searching new kinds of interactions between barions (hypothetical fifth force~\cite{Ubachs2013} or corrections to gravitational interactions at the nanometer scale~\cite{Borkowski2016}). The relatively small magnitude of the van der Waals forces in these molecules will yield better constraints on such exotic forces.

In view of the above advantages, even a single-species mercury MOT seems to be very interesting,  yet remains challenging to produce experimentally. To date, only five groups have reported successful realization of the mercury MOT \cite{PhysRevLett.100.053001,Villwock2011,Yi2011,Hong-Li2013,Paul2015}. Still, quantum degeneracy has not been accomplished and details of ultra cold collisions in mercury remain unknown.

In Sec. 2 we describe in details the experimental set-up of a double-species MOT for various isotopes of Hg and $^{87}$Rb. In Sec. 3 we report preliminary experimental results. We conclude the article in Sec 4.

\section{Experimental set-up} 

A MOT for trapping and cooling of two different species requires two independent laser systems and a vacuum set-up with a common science chamber. Rubidium and mercury atoms are loaded into the science chamber by a Zeeman slower and through a 2D-MOT chamber, respectively. The most essential elements of the experimental set-up are described below.

\subsection{Vacuum system}

\begin{figure}
  \centering\includegraphics[width=0.9\textwidth]{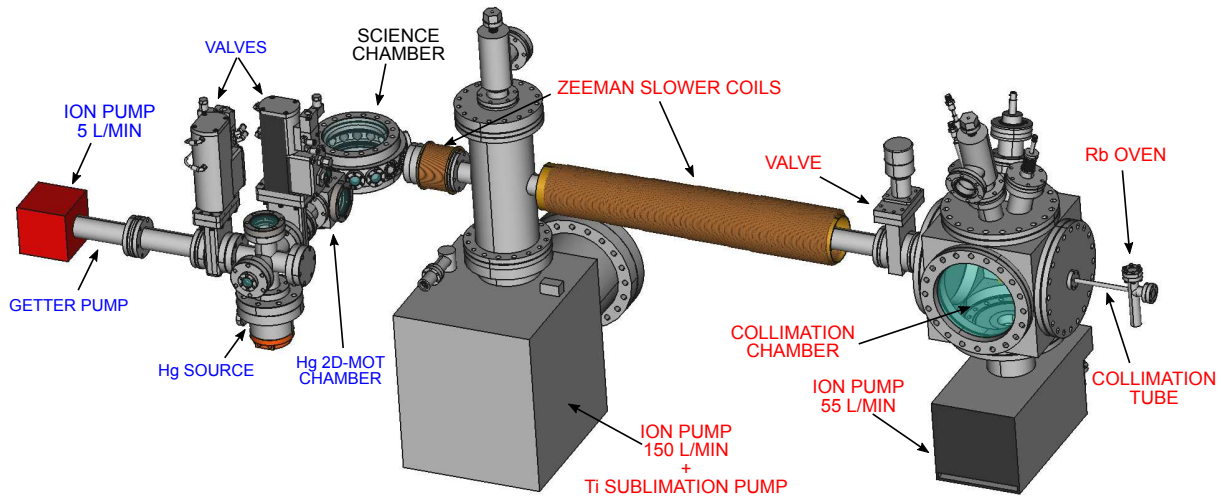}
  \caption{\label{vac}Schematic of the vacuum assembly. Rubidium oven is heated up to 120$^\circ$C to provide high enough flux of atoms. Mercury source chamber contains a drop of mercury which is kept in the temperature of -8$^\circ$C during experiment (-30$^\circ$C in the idle state). Red and blue labels correspond to Rb and Hg part, respectively. }
\end{figure}

A 3D-view of the vacuum assembly, used for cooling and trapping of Rb and Hg atoms, is shown in Fig.~\ref{vac}. The system consists of two perpendicular arms connected to the science chamber.  These arms, shorter and longer, provide beams of mercury and rubidium atoms, respectively.

The project of the rubidium vacuum part is an improved version of the Bose-Einstein condensation set-up developed by the group of Aspect~\cite{Desruelle1998, Desruelle1999}. It consists of a rubidium source, a collimating chamber and a Zeeman slower.

The rubidium atomic beam is produced by a recirculating oven~\cite{Swenumson1981, Lambropoulos1977}. The oven is based on a concept presented elsewhere~\cite{fauquembergue}. In the oven a natural isotopic sample of rubidium is heated in a reservoir up to 120$^\circ$C. The atomic beam is released from the heated oven through a long (10~cm) and thin (5~mm) collimation tube. To increase the temperature of the sample we use two heating coils  wound around both reservoir and collimation tube. Under typical experimental conditions it provides a flux of  5$\times$10$^{11}$~atoms/s. Rubidium atoms can stick to the inner walls of the thin tube and block it.  To avoid this effect a temperature gradient between 100$^\circ$C and 40$^\circ$C along the collimation tube is maintained. Additionally, the inner wall of the collimation tube is covered by a thin layer of a stainless-steel mesh. The temperature gradient combined with the presence of the mesh ease the recycling of rubidium back to the reservoir.  
The collimation tube of the oven is connected directly to the collimation chamber.  Inside the chamber the outgoing atomic beam can be blocked by an in-vacuum shutter made of copper.  Subsequently, rubidium atoms pass through a 5~mm aperture in a cold copper block.  The collimation chamber is pumped by a single 55~l/min noble diode ion pump (Varian) to the pressure of 10$^{-8}$~mbar. To improve the vacuum conditions the block is chilled to -20$^\circ$C by a flow of a cooling liquid. 

\begin{figure}
  \centering\includegraphics[width=0.5\textwidth]{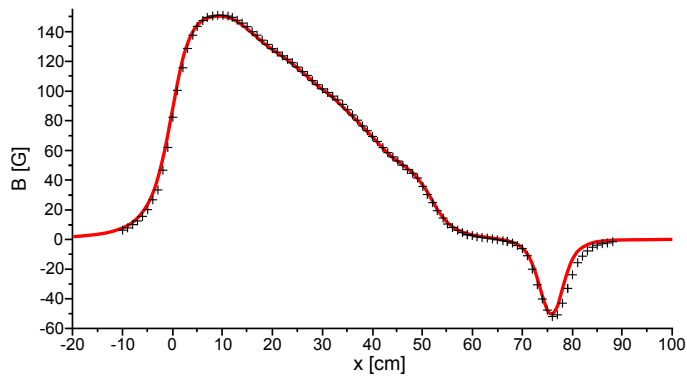}
  \caption{\label{ZS_field}Magnetic field along the Zeeman slower 
  axis. Solid curve shows the profile optimized for $^{87}$Rb atoms. Cross-points show the 
experimental data.}
 \end{figure}

After the collimation chamber the atomic beam passes through a gate valve and enters the Zeeman slower \cite{Phillips1982}. The calculated and measured Zeeman slower field profiles are depicted in Fig.~\ref{ZS_field}.  A spin-flip slower design \cite{Courtillot2003} with two multilayer conical coils is used. This solution has an advantage of operating at low currents (typically of 5~A and 0.35~A in the bigger and smaller coil, respectively) with no requirements of water cooling.

\begin{figure}
   \centering\includegraphics[width=0.4\textwidth]{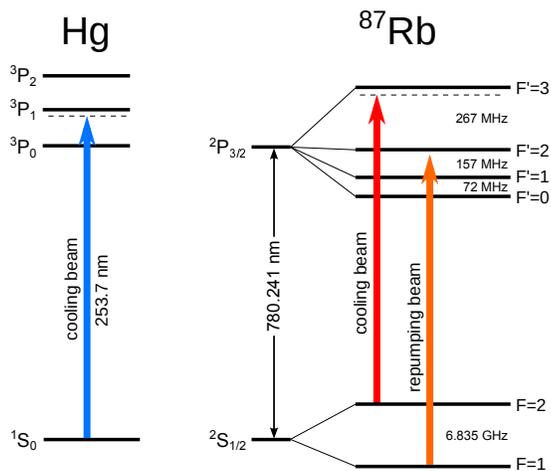}
   \caption{\label{levels}Atomic energy levels and transitions used for magneto-optical trapping and cooling of bosonic Hg and $^{87}$Rb atoms.  }
 \end{figure}

 The atoms in the Zeeman slower are slowed by a 14~mW circularly polarized laser beam. The beam is counter-propagating with respect to the rubidium atomic beam and is focused on the output of the oven. The frequency of the laser beam is detuned -109~MHz from the $|^2$S$_{1/2}$~F=2$\rangle \rightarrow$$|^2$P$_{3/2}$~F=3$\rangle$ resonance in $^{87}$Rb. To maintain atoms in the cycling transition a repumping beam is used, Fig.~\ref{levels}. The 3~mW repumping beam is circularly polarized and is detuned +123~MHz from the $|^2$S$_{1/2}$~F=1$\rangle \rightarrow$$|^2$P$_{3/2}$~F=2$\rangle$ resonance. The cooling and the repumping beams copropagate. 

The most probable velocity of the rubidium atoms entering the slower is about 330~m/s. Under typical experimental conditions our Zeeman slower is able to cool about 30\% of $^{87}$Rb atoms from the corresponding velocity distribution.  The final velocity of slowed rubidium atoms at the output of the Zeeman slower is about 40~m/s which is low enough to capture them efficiently inside the MOT. 

The MOT itself is created in the hexadecagon stainless steel science chamber (Kimball Physics). Its two CF100 windows and fourteen CF16 windows are accessible in vertical and horizontal directions, respectively, as shown in Figs.~\ref{vac} and \ref{beams}. Another two CF16 viewports connect the science chamber to both arms of the vacuum system.

The mercury part of the vacuum system consists of a mercury source, a pumping stage and a 2D-MOT chamber. 
The realization of the source chamber for mercury is based on a project of the Bize group \cite{petersen,Yi2011}. Mercury has a high vapour pressure at room temperature and does not require an oven to produce the gas state.  On the contrary, it needs to be cooled substantially to maintain high-vacuum conditions. To lower the temperature of the mercury a two-stage thermoelectric cooler (TEC) is used. The cold side of the cooler is connected to a copper container in which a small drop of mercury is placed. The hot side of the thermoelectric element is connected to a water cooled copper heat sink. The temperature difference between top and bottom of the cooler is up to 50$^\circ$C which results in a -30$^\circ$C of the mercury drop and the vapour pressure of about 5$\times$10$^{-8}$~mbar. The temperature of the mercury drop, and in consequence mercury vapour pressure, is regulated by adjusting TEC current. The typical temperature used in experiment is set to -8$^\circ$C.

The mercury source chamber is pumped by a combination of getter and ion pumps (SEAS Getters NEXTorr D200-5). To extend the lifetime of the pumps the source chamber can be remotely isolated from the rest of the vacuum system by two pneumatic valves, e.g.  in emergency situations when cooling of mercury is impossible.

Between the science and the source chambers a two-dimensional magneto-optical trap (2D-HgMOT) is placed. It is meant to be used as a preselecting stage of slower mercury atoms before loading them into the MOT. 
 2D-MOT is not operating at the current stage of the set-up, since we found loading the 3D-MOT directly from the mercury background vapour efficient enough. Nevertheless, it will improve the loading rate substantially in future applications~\cite{petersen,Dieckmann1998}. 

To maintain low pressure in the science chamber, differential pumping schemes were implemented for rubidium and mercury sources. The science chamber is pumped with a 150~l/min ion pump and a titanium sublimation pump combined in one enclosure (Varian). The differential pumping between the science chamber and the rubidium oven is realized by a 50~cm long graphite tube with internal diameter of 15~mm.  The graphite tube, which adsorbs rubidium atoms, is placed inside the Zeeman slower.

Another differential pumping scheme is implemented between the science and 2D-HgMOT chambers.  They are connected by a 47~mm long copper cylinder with drilled tube. The internal diameter of the tube is equal to 1~mm at the input side. After 25~mm its diameter starts to expand up to 15~mm at the output side. This conically shaped part of the tube is coated by a thin layer of gold.  The surface was grooved before coating to increase an active area.  Mercury atoms stick to the golden layer and create an amalgam.  It improves pumping efficiency and  allows the vacuum pressure in the science chamber to reach 10$^{-9}$~mbar.

 \subsection{Laser system} 

 \begin{figure}
   \centering\includegraphics[width=\textwidth]{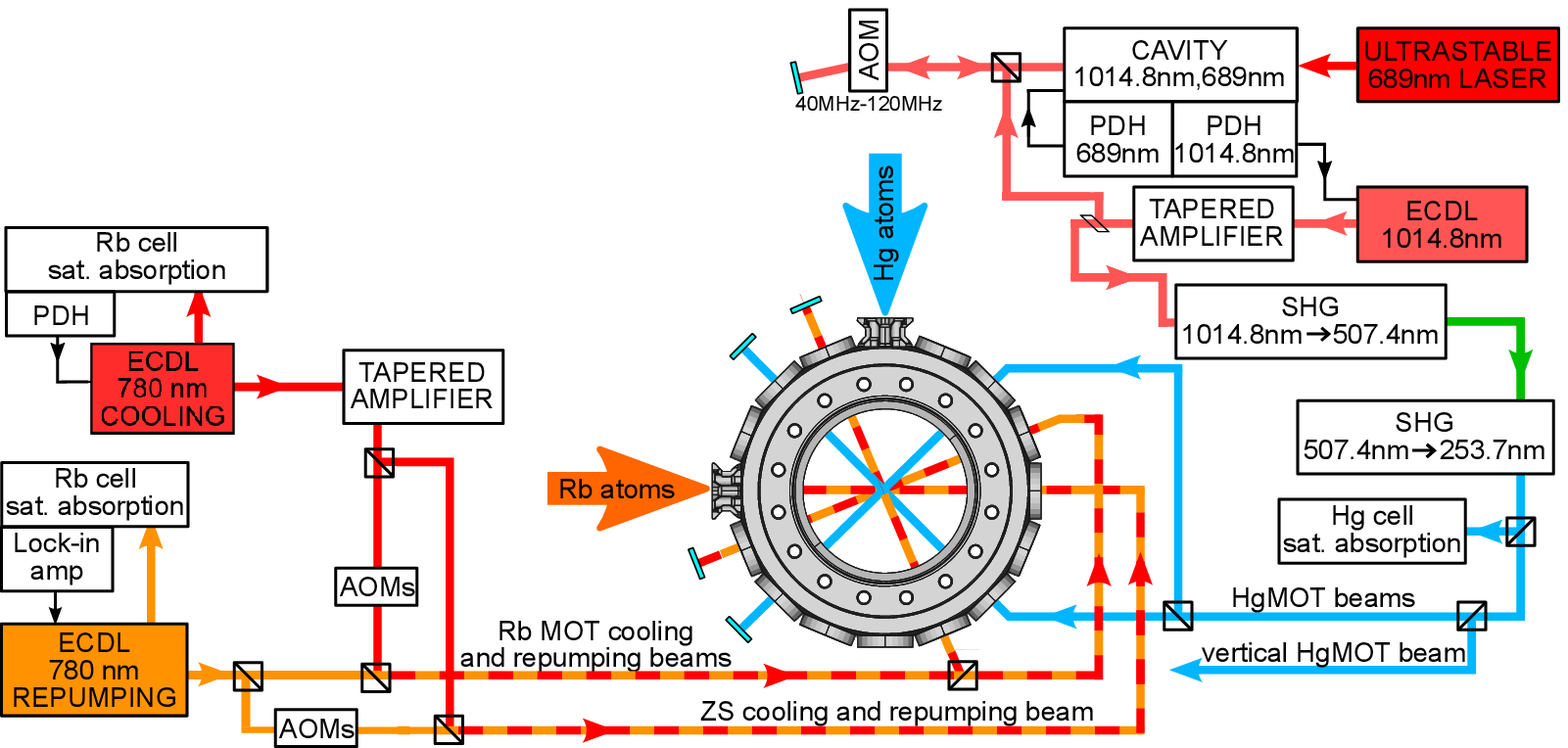}
   \caption{\label{beams}A simplified scheme of the laser system for the $^{87}$Rb-Hg MOT. The cooling and horizontal repumping beams for the Rb MOT as well as the cooling and repumping beams for Zeeman slowing are combined  on polarizing beam splitters (PBS). Both Rb lasers are stabilized to atomic transitions with saturated absorption spectroscopy. The 1014.8~nm laser is stabilized to the 689~nm ultra-stable laser through the transfer cavity. Frequency of the beams in different parts of the set-up are controlled by acousto-optic modulators (AOM). 
 }
 \end{figure}

 Separate laser systems are used for cooling and trapping of rubidium and mercury, Fig.~\ref{beams}. Diagrams of appropriate energy levels used in the cooling process for both species are shown in Fig.~\ref{levels}. 

 Rubidium requires separate lasers for cooling and repumping light as the value of the ground state hyperfine splitting is large.  The cooling and repumping lasers are commercial external-cavity diode lasers (ECDL) made by Toptica Photonics. The cooling laser is frequency stabilized by the Pound-Drever-Hall (PDH) locking scheme \cite{Drever1983} to an atomic resonance in a rubidium vapour cell. A small portion of the light from the cooling laser is frequency shifted by -68~MHz by an acousto-optic modulator (AOM) and then is directed to the saturated spectroscopy set-up.  The laser frequency is locked to the crossover resonance between $|^2$S$_{1/2}$ F=2$\rangle \rightarrow$$|^2$P$_{3/2}$ F=2$\rangle$ and $|^2$S$_{1/2}$ F=2$\rangle \rightarrow$$|^2$P$_{3/2}$ F=3$\rangle$ transitions in $^{87}$Rb. 

 The cooling laser beam is amplified by a fibre-coupled tapered amplifier (Thorlabs) up to 700~mW and then passes through a single-mode polarization-maintaining fibre. About 35~mW of the beam is decoupled, upshifted +92~MHz by an AOM and sent to the Zeeman slower. The rest of the cooling laser light is shifted by +188.8~MHz using an AOM in a double-pass configuration to be red-detuned by about two natural linewidths (2$\Gamma_{Rb}$=12.2~MHz) from the $|^2$S$_{1/2}$ F=2$\rangle \rightarrow$$|^2$P$_{3/2}$ F=3$\rangle$ transition in $^{87}$Rb.
 The laser light is expanded to a 1/e$^2$ diameter of 15~mm, then split into three beams and sent to the science chamber. The beams are circularly polarized and have the same power of 17~mW which results in a laser intensity of 5.8$I_{sat}$, where $I_{sat}$ stands for the saturation intensity of the cooling transition in Rb atoms. To maximize the power available for cooling a retro-reflection configuration is implemented. 
 Two of the cooling beams enter the science chamber along the horizontal perpendicular directions. The third one is deviated by 17$^\circ$ from the vertical direction to avoid overlapping with one of the Hg-MOT trapping beams.

 The repumping laser frequency is stabilized to the $|^2$S$_{1/2}$ F=1$\rangle \rightarrow$$|^2$P$_{3/2}$ F=2$\rangle$ transition in $^{87}$Rb using saturated absorption signal. The length of the extended cavity of the laser is modulated with the frequency of 5.5~kHz by changing the length of a PZT element. The error signal is generated by a lock-in amplifier (Stanford Research Systems SR830). A part of the repumping beam (5~mW) is expanded to a diameter of 12~mm and combined with two horizontal cooling beams.  Another part of the repumping beam is +123~MHz frequency shifted by two AOMs and then superimposed with the Zeeman slower cooling beam.

 Cooling and trapping of mercury atoms requires a UV-C laser source at 253.7~nm to drive the $|^1$S$_0\rangle$$\rightarrow$$|^3$P$_1\rangle$ intercombination transition. 
 Due to the lack of commercially available powerful and tunable lasers working directly in the UV-C range, a 1014.8~nm  laser is used as a source and its light is frequency-doubled twice in a system of two doubling stages (Leos Company). The source laser is composed of a home-made ECDL with the output power of 50~mW and a tapered amplifier (Sacher Lasertechnic) capable of generating an output power of 1.1~W. Power build-up bow-tie cavities are used to improve the efficiency of frequency doubling.  
 Both cavities are locked to the input light frequency by the H\"ansch-Couillaud method~\cite{Hansch1980}. In the first cavity (1014.8~nm$\rightarrow$507.4~nm) a lithium triborate (LBO) crystal is used as the nonlinear medium whilst in the second one (507.4~nm$\rightarrow$253.7~nm) a $\beta$-barium borate (BBO) crystal is used. Both crystals are heated to 50$^\circ$C to achieve optimal phase matching. The conversion efficiencies are 48\% and 13\% for the first and the second cavity, respectively. 
 Up to 60~mW of UV power is generated by the quadrupling system. Almost all the UV power is sent to the Hg MOT, while a small portion (2~mW) is sent to a 1~mm-thick Hg-vapour reference cell.  The saturation spectroscopy enables pre-tuning the UV radiation frequency near the atomic resonance for a given mercury isotope.  The MOT beam is expanded through a telescope to have a 1/e$^2$ diameter of 6~mm. The beam is then separated into three circularly polarized beams which operate in strictly perpendicular retro-reflection scheme.

 The UV laser frequency is stabilized to the narrow (7.5~kHz) $|^1$S$_{0}\rangle\rightarrow$ $|^3$P$_1\rangle$ transition in Sr by a stability transfer from an ultra-stable laser\cite{Lisak2012,Cygan2013}. The ultra-stable laser is a part of an optical strontium lattice clock~\cite{Morzynski2015}. A small part of the 1014.8~nm light is uncoupled from the main beam before quadrupling and provided through an AOM and a single-mode polarization-maintaining fibre to a transfer cavity with finesse equal to 150 and 50 for the wavelength of 689~nm and 1014.8~nm, respectively.  The length of the transfer cavity is stabilized to the 689~nm ultra-stable laser by the PDH method. The same method is used to lock the fundamental 1014.8~nm laser to the transfer cavity mode.  This approach enables control of the frequency at a level better than the natural linewidth of the mercury cooling transition ($\Gamma_{Hg}=$1.27~MHz). 

 Prior to the transfer cavity the fundamental laser beam double-passes AOM which shifts the frequency of the beam in the range between +80~MHz and +240~MHz. This way, taking advantage of the frequency quadrupling, we are able to  scan the UV laser over a range of 640~MHz while the fundamental frequency of the 1014.8~nm laser remains locked.  Since no UV light is used for locking, almost all of the UV power is provided into the MOT which is a big advantage of the method. 

 In our experimental set-up several mercury isotopes can be cooled and trapped. The free spectral range of the transfer cavity equal to 1~GHz combined with the UV laser tuning range allows fast and easy switching between isotopes.  

 \subsection{MOT}

 The MOT for both species is created in the middle of the science chamber.  The laser beams of the Zeeman slower do not affect the atoms in the  MOT, since their frequencies are far from any Hg or Rb resonance.

 To produce the trapping magnetic field a pair of coils in an anti-Helmholtz configuration is used. The coils are placed externally to the chamber and provide quadrupole magnetic field with an axial gradient of 15~G/cm at the centre of the trapping area while operating at 8.5~A. To prevent excessive heating the coils are water cooled.  Stray magnetic fields are compensated by a system of three additional pairs of coils.  This configuration allows achieving up to 10$^7$ rubidium atoms and up to 2$\times$10$^6$ mercury atoms in the MOT while operating in one species mode. However, to overlap both species they need to be shifted from their optimal positions, which results in a decrease of the number of atoms in both clouds. The typical number of atoms achieved in double species experiment is of the order of 10$^5$ in each cloud. 

 The number of atoms in the MOT is determined by fluorescence imaging on a CCD camera (Apogee U47-HC). The camera operates at both wavelengths 780~nm and 253.7~nm with 57\% and 70\%  quantum efficiency, respectively. The fluorescence light from the MOT can be recorded along two orthogonal directions such that precise characterization of spatial overlap of both atomic clouds is possible. The clouds can be imaged either together or separately. The observed wavelength is chosen by a rotatable holder with appropriate interference filters placed near the camera sensor. 

\begin{figure}
   \centering\includegraphics[width=0.9\textwidth]{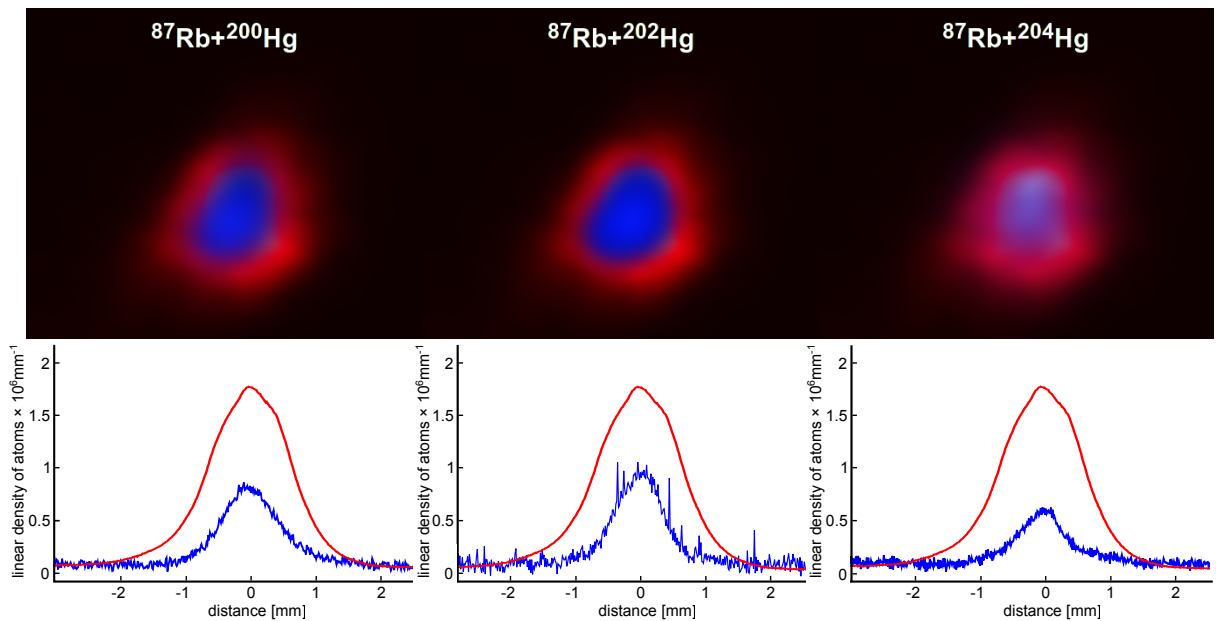}
   \caption{\label{mots}Upper: false colour fluorescence images of the rubidium (red) and mercury (blue) MOTs. Lower: Linear density of atoms in rubidium (red) and mercury (blue) MOTs. The profiles are the result of integration of the number of atoms along the imaging direction. The profiles correspond to the images of MOTs. }
 \end{figure}

 Since the retro-reflection configuration of cooling and trapping beams is used, the return beams are less intense which results in a radiation pressure imbalance and shifts the position of the MOTs away from the zero magnetic field. We observe that after individual alignment the atomic clouds do not overlap completely. To achieve good spatial overlap more careful adjustment is needed. Typical images taken for  overlapped  rubidium and mercury MOTs are shown in Fig.~\ref{mots}.  

 The characterization of the MOT can be performed either by analysis of fluorescence images of the MOT taken by the CCD camera or by measuring the fluorescence light from the MOT by a photomultiplier tube (PMT, Hamamatsu C659-S). Additionally, fluorescence light from the rubidium MOT is focused on a calibrated photodiode.

 A standard time of flight absorption measurement technique is used to measure the temperature of the rubidium MOT. The temperature is below 160~$\mu$K under our typical experimental conditions, i.e. light power of 17~mW in each of the trapping beams and frequency detuning of two natural linewidths. The temperature of the mercury MOT is estimated to be below 100~$\mu$K by the release and recapture technique. The estimation is performed for the total power of 23~mW of the cooling light and frequency detuning of 3.5 natural linewidths. 

 \section{Results}
 \subsection{Loading of the MOT} 
 \label{ssub:Loading of the MOT}

 Loading dynamics of a single species MOT can be described by the following rate equation:

 \begin{equation}\label{eq:loading-n}
   \frac{dN(t)}{dt}=R-\beta N(t) - \chi \int n^2(t) dV,
 \end{equation}

\noindent
  where $N(t)$ and $n(t)$ are the number and density of atoms in the trap, respectively, and $R$ is the rate at which atoms are collected in the MOT. The trap loss rate coefficients, $\beta$ and $\chi$, describe collisions between cold atoms and hot background atoms, and two-atoms cold collisions, respectively.
If we assume that initially the density of the trapped atoms in the cloud increases till reaching their maximum value in the centre (temperature limit)~\cite{Weiner1999}, we can simplify Eq.~(\ref{eq:loading-n}) into 

 \begin{equation}\label{eq:loading-N}
   \frac{dN(t)}{dt}=R-\beta N(t) - \gamma N^2(t),
 \end{equation}
 where $\gamma \propto \chi/V$.
\noindent
The solution of Eq.~(\ref{eq:loading-N}),  considering boundary conditions for loading process, i.e $N(t=0)=0$ and $N(t\rightarrow\infty)=N_{s}$ where $N_{s}$ describes the number of atoms in stationary state, is given by

 \begin{equation}\label{eq:loading_solution}
   N(t) = N_s \frac{(\eta+\beta)\left(1-\exp(\eta t)\right)}{(-\eta-\beta)\left(1+\exp(\eta t)\right)+2\beta},
 \end{equation}

\noindent
where $\eta = (\beta^2 + 4 R \gamma)^{{}^{1}/{}_{2}}$ and $N_S = (\eta-\beta)/(2\gamma)$. If the two-body losses can be neglected, Eq.~(\ref{eq:loading-N}) can be simplified even further:

 \begin{equation}\label{eq:loading}
   \frac{dN(t)}{dt}=R-\beta N(t).
 \end{equation}
The solution of the Eq.~(\ref{eq:loading}) yields an exponential growth in the number of atoms in the MOT:

 \begin{equation}\label{eq:number_of_atoms}
   N(t)=N_s[1- \exp(-\beta t)],
 \end{equation}

\noindent
where $N_s=R/\beta$ is a steady state number of atoms.

\begin{figure}
   \centering\includegraphics[width=0.85\textwidth]{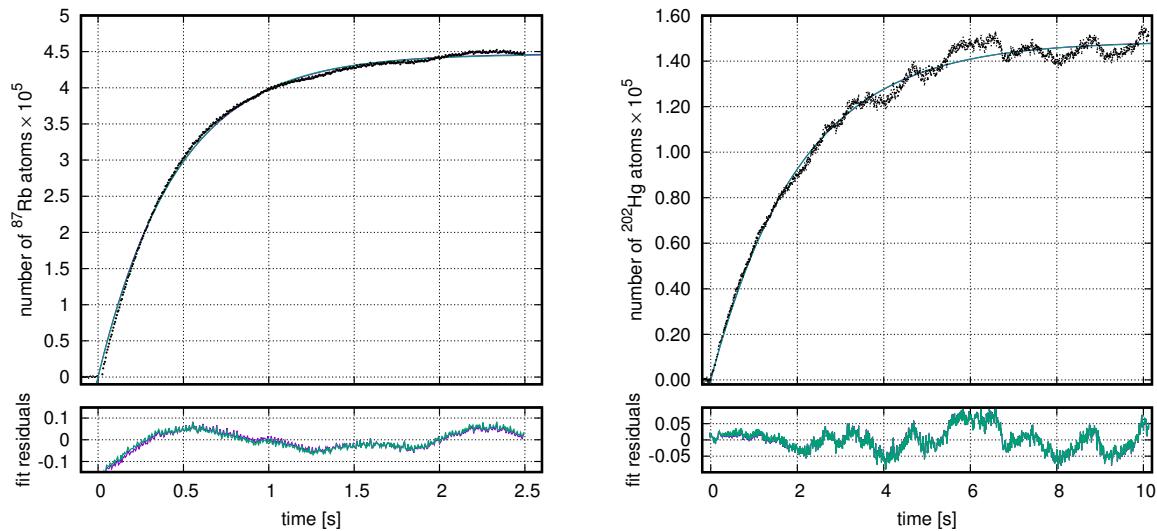}
   \caption{\label{loading_curves_hg_rb}Upper panels: loading curves of the $^{87}$Rb (left) and $^{202}$Hg (right) single component MOT. The measurements were performed under our typical experimental conditions. The frequency detuning and the light power was equal to two (three) natural linewidths and 17~mW (3~mW) in each of the $^{87}$Rb ($^{202}$Hg) trapping beams, respectively. The violet and green curves (indistinguishable in the upper panels) are fits to experimental data of the Eq.~(\ref{eq:loading_solution}) and Eq.~(\ref{eq:number_of_atoms}), respectively. Lower panels: residuals of the fits.}
 \end{figure}

 Fig.~\ref{loading_curves_hg_rb} presents loading curves of the $^{87}$Rb and $^{202}$Hg single component MOT recorded under our typical experimental conditions. The functions \ref{eq:loading_solution} and \ref{eq:number_of_atoms} were fitted to experimental data. The lower panels of Fig.~\ref{loading_curves_hg_rb} depicts the residuals of the fit. In both $^{87}$Rb and $^{202}$Hg  single component MOTs in our experimental conditions we do not observe light-assisted homonuclear collisions, i.e. fit of Eq.~(\ref{eq:loading_solution}) yields $\eta \simeq\beta$. The fitted values of the $1/e$ loading times, i.e. $1/\beta$, are equal to  0.44~s and 2.1~s for the $^{87}$Rb-MOT and $^{202}$Hg-MOT, respectively. 
 
 \subsection{Photoionization effect in the $^{87}$Rb-MOT}

 \begin{figure}
   \centering\includegraphics[width=0.6\textwidth]{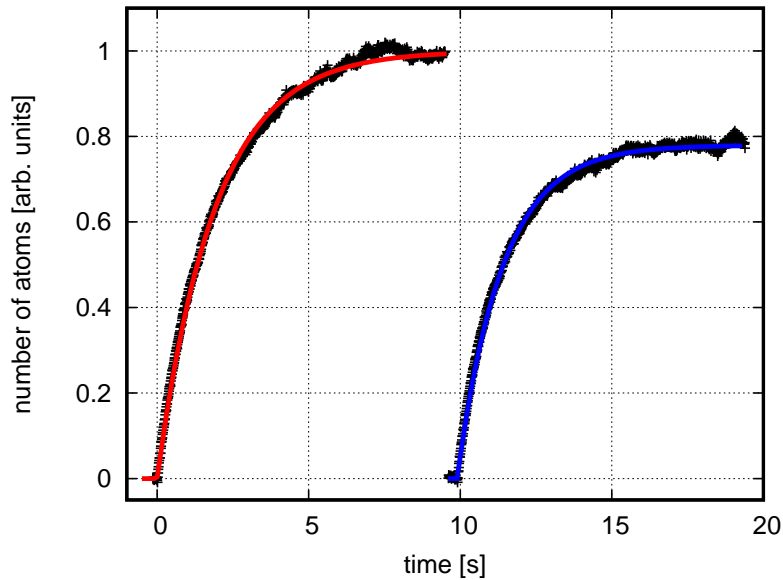}
   \caption{\label{photoionization}Successively measured loading curves of the $^{87}$Rb-MOT. The first curve was detected in the absence of the UV light while the second curve was measured in the presence of the UV light (20 mW). The red and blue curves are exponential fits to the Eq.~(\ref{eq:number_of_atoms}) and Eq.~(\ref{eq:number_of_atoms_modified}), respectively. The reduction of the number of rubidium atoms is clearly seen, however, simultaneous trapping of rubidium and mercury is possible even with relatively high intensity of the UV light.}
 \end{figure}
 The rubidium atoms are exposed to the radiation of the 253.7~nm UV light while the double-species set-up is operating. The energy of UV photons is high enough to ionize rubidium atoms in both ground (5$^2$S$_{1/2}$) and excited (5$^2$P$_{3/2}$) states since photoionization from these states requires a photon energy of 297 nm and 479 nm, respectively. Thus, the photoionization process results in losses of rubidium atoms from the MOT. Another term $\beta_p$ needs to be introduced to the Eq.~(\ref{eq:loading}) to describe these additional losses. It yields the following form of the rate equation:

 \begin{equation}\label{eq:loading_photoionization}
   \frac{dN}{dt}=R-(\beta+\beta_p) N.
 \end{equation}
 
The photoionization rate $\beta_p^{\left|X\right\rangle }$ for atoms in MOT in a given state ${\left|X\right\rangle }$  can be derived from a simple formula  

 \begin{equation}\label{eq:rate_at_X}
   \beta_p^{\left|X\right\rangle } = \sigma^{\left|X\right\rangle }(\nu) \frac{I_p}{h\nu} f^{\left|X\right\rangle },
 \end{equation}
 
 \noindent
  where  $I_p$  is the
  effective average intensity of the photoionization beam seen by the atoms in the MOT, $h\nu$ is the energy of the ionizing photon, $\sigma^{\left|X\right\rangle }(\nu)$ is the photoionization cross section at the $\nu$ frequency, and $f^{\left|X\right\rangle }$ is a fraction of atoms in this state. Note that the $I_p$ differs from the intensity of the photoionizing beams, since generally the size of the MOT is smaller than the size of the photoionizing beams.  The $I_p$ seen by the atoms can be derived by averaging the intensity distribution of the photoionizing beam over the spatial distribution of the
atoms \cite{Lowell2002}.
 
   The theoretical estimation of the photoionization cross section for the 5S${}_{1/2}$ ground state in Rb at $253.7$~nm, derived from the semiempirical theory  \cite{Weisheit1972}, is in the range of $(4.5-7.2)\times 10^{-21}$cm${}^2$.
    The excited-state 5P$_{3/2}$ photoionization cross section  of Rb atoms calculated at 253.7~nm using quantum defect theory  \cite{Aymar1984} is in the range of $(1-10)\times 10^{-18}$cm${}^2$ (inferred from plot) which is three orders of magnitude greater than in the ground state.
 
    If we assume that the photoionization losses can be calculated separately for atoms in each of the states the theoretical  photoionization rates  for our experimental conditions can be deduced if the populations of the 5S${}_{1/2}$ and 5P${}_{3/2}$ states in the Rb-MOT are known. A typical total (sum of all six UV beams at $\lambda$=253.7~nm, taking into account power transmissions of vacuum windows) intensity used in our dual-species experiments is equal  to 57~mW/cm$^2$. The geometrical sizes of UV beams and Rb-MOT yields the effective average intensity  equal to 49~mW/cm$^2$

 To quantify the populations of the 5S${}_{1/2}$ and 5P${}_{3/2}$ states in the Rb-MOT we performed photoionization measurements of the 5P${}_{3/2}$ state with an additional 401~nm diode laser. We measured  dependence of the photoionization rate of the 5P${}_{3/2}$ state on the Rb-MOT cooling light intensity. A fit of the saturation curve \cite{Dinneen1992,Javanainen1993} yields the fractions of atoms in the S${}_{1/2}$ and 5P${}_{3/2}$ states equal to 0.73(2) and 0.27(2), respectively, in our typical experimental conditions. 
 
  The values written above, i.e. cross sections, effective intensity and fractions of atoms, yield the theoretical values of the 253.7~nm photoionization rates for the 5S${}_{1/2}$ and 5P${}_{3/2}$ states equal to $1.65(63)\times10^{-4}$~s$^{-1}$ and 
 $0.085(76)$~s$^{-1}$, respectively .

  To directly determine the photoionization rate $\beta_p$ in our experiment   the loading curves of the Rb-MOT were analyzed.
 
 Solving the Eq.~(\ref{eq:loading_photoionization}) time dependence of the number of atoms can be written as:

 \begin{equation}\label{eq:number_of_atoms_modified}
   N(t)=N'_s[1-exp(-(\beta+\beta_p) t)].
 \end{equation}

 The steady state number of rubidium atoms irradiated by the UV light is reduced to $N^{'}_{s}=R/(\beta+\beta_p)$. 
 A difference between the loading rates of two successively measured loading curves with and without the presence of the UV light, respectively, yields the photoionization rate $\beta_p$, Fig.~\ref{photoionization}. 

  The resulting value of the photoionization rate is equal to 0.151(9)~s$^{-1}$. The photoionization rate measured in our system is close to the value deduced above for the 5P${}_{3/2}$ state. Nevertheless, a complete theory describing the system where two states which can be ionized are strongly coupled by the Rb trapping light beams should include possible  interferences \cite{Takekoshi2004,Yin1992}.
   
 \subsection{Dual-species Hg-Rb MOT}

\begin{figure}
   \centering\includegraphics[width=0.6\textwidth]{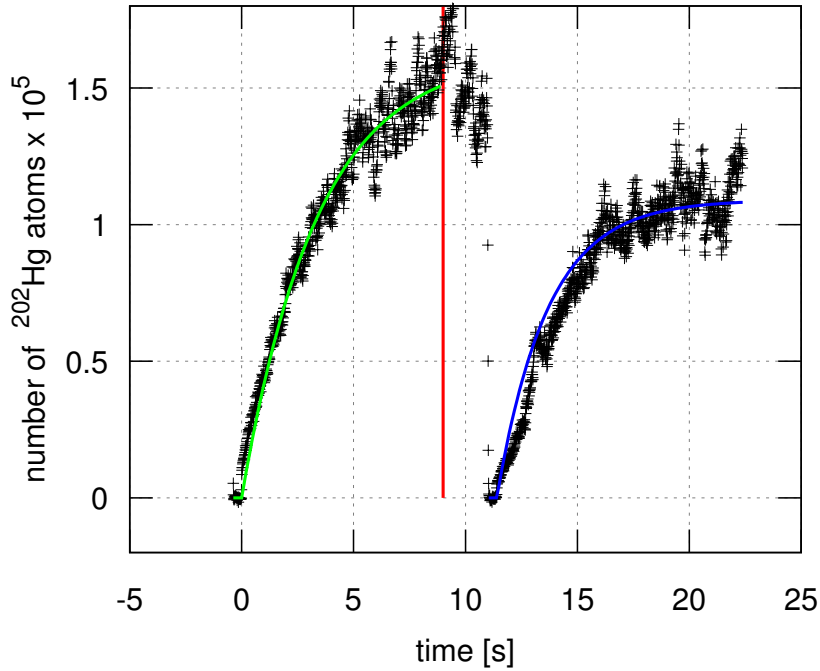}
   \caption{\label{hg-rb_interact}Loading curves of the $^{202}$Hg-MOT. Firstly, the $^{202}$Hg-MOT is loaded in the absence of the trapped rubidium atoms, then the trap is subsequently depleted and loaded again in the presence of the $^{87}$Rb trapped in the MOT. The vertical red line denotes the moment at which rubidium atoms start to be loaded into the MOT.  The steady state number of Rb atoms is 8~times larger than the number of Hg atoms while both species are loaded simultaneously into the MOT. The blue and green curves are exponential fits to the Eq.~(\ref{eq:loading_Hg_simp_time_dep}) and Eq.~(\ref{eq:loading_Hg_simp_time_dep_no_rb}), respectively.}
 \end{figure}

  The loading dynamics of the $^{202}$Hg-MOT was examined with and without its rubidium counterpart. Fig.~\ref{hg-rb_interact} shows the comparison of two loading curves of $^{202}$Hg-MOT. First curve (green) was measured in the absence of rubidium atoms when the cooling and repumping beams for rubidium atoms were switched off. Subsequently, the rubidium MOT beams were switched on and $^{87}$Rb-MOT was completely loaded. Then the frequency of the UV trapping light was tuned to be on the blue side of the $^{202}$Hg cooling transition for a few miliseconds which results in emptying of the $^{202}$Hg-MOT. Finally, the second loading curve of the $^{202}$Hg-MOT  (blue) was measured in the presence of the trapped $^{87}$Rb atoms.   

  The loading of the Hg-MOT in the presence of the Rb-MOT can be modelled by the modified rate Eq.~(\ref{eq:loading}):
\begin{equation}\label{eq:loading_Hg}
      \frac{dN_{Hg}}{dt}=R_{Hg}-\beta_{Hg}N_{Hg} -\gamma_{Hg,Rb}\int n_{Rb}n_{Hg}dV,
    \end{equation}
where 
$\gamma_{Hg,Rb}$ is the two-body loss rate accounting for the losses of Hg atoms because of collisions with Rb atoms captured in the same MOT.
The steady state number of atoms in the Hg-MOT is equal to 1.62 $\times$ 10$^5$, whilst the corresponding number of Rb atoms is 8~times larger. 

The difference in the number of atoms gives a simplification: $\int n_{Hg} n_{Rb} dV= n_{Rb}N_{Hg}$ allowing the Eq.~(\ref{eq:loading_Hg}) to be written as:

\begin{equation}\label{eq:loading_Hg_simp}
  \frac{dN_{Hg}}{dt}=R_{Hg}-(\beta_{Hg}+\gamma_{Hg,Rb}n_{Rb})N_{Hg},
\end{equation}
which leads to the following time dependence of the number of Hg atoms while loading in the presence of the Rb-MOT:
\begin{equation}\label{eq:loading_Hg_simp_time_dep}
  N_{Hg}(t)=\frac{R_{Hg}}{B_{Hg,Rb}}[1-exp(-B_{Hg,Rb}t)],
\end{equation}
where $B_{Hg,Rb}=\beta_{Hg}+\gamma_{Hg,Rb}n_{Rb}$. 

In the absence of Rb atoms the Eq.~(\ref{eq:loading_Hg_simp_time_dep}) can be rewritten as:
\begin{equation}\label{eq:loading_Hg_simp_time_dep_no_rb}
  N_{Hg}(t)=\frac{R_{Hg}}{\beta_{Hg}}[1-exp(-\beta_{Hg}t)].
\end{equation}

The value of the $\gamma_{Hg,Rb}$ was estimated on the basis of the equation $\gamma_{Hg,Rb}=(B_{Hg,Rb}-\beta_{Hg})/n_{Rb}$ to be of the order of $10^{-9}$~cm$^3$s$^{-1}$. The values of $\beta_{Hg}=0.297(34)$~s$^{-1}$ and $R_{Hg}=0.0207(22)$~s$^{-1}$ ($B_{Hg,Rb}=0.442(16)$~s$^{-1}$) were obtained from the fit of Eq.~(\ref{eq:loading_Hg_simp_time_dep_no_rb}) (Eq.~(\ref{eq:loading_Hg_simp_time_dep})) to the experimental data presented in Fig.~\ref{hg-rb_interact}. The density of the Rb-MOT was determined to be 1.03(10)$\times$ 10$^8$~cm$^{-3}$ by analysis of the CCD picture of the Rb-MOT.

We checked that in our system the Rb-MOT beams do not have any impact on the number of trapped mercury atoms. Therefore, our measurements indeed show existence of interspecies collision-induced losses in the $^{87}$Rb-$^{202}$Hg system. 

Regular fluctuations are visible in the loading curve of the Hg-MOT in the presence of the Rb-MOT. They are most clearly noticeable in the steady state, but they also disturb the initial part of the loading curve yielding a discrepancy between experimental data and the exponential fit (blue curve in Fig. 8). The origin of these fluctuations is not clear, nevertheless it seems to be related to the presence of cold Rb atoms, since such fluctuations are not evident in the loading of the single component Hg MOT.

Describing losses of Rb atoms due to collisions with Hg atoms, i.e. $\gamma_{Rb,Hg}$ value, is more challenging, since losses due to photoinization have to be taken into account. Moreover, the number of Hg atoms is lower by about one order of magnitude than Rb atoms. Therefore, the impact of Hg atoms on Rb-MOT is significantly smaller  compared to the opposite situation described above. At this point we are not able to observe the influence of Hg atoms on the number of atoms in the Rb-MOT.

 \section{Conclusion} 
 \label{sec:Conclusion}

 We have demonstrated simultaneous trapping of rubidium and mercury atoms in the same MOT. This novel experimental set-up paves the way for further exciting experiments, like the photoassociation of rubidium and mercury atoms.  We have investigated the unique challenges of this set-up: the photoionization of the rubidium atoms by the lasers used to cool the mercury atoms and loss of mercury atoms due to the presence of cold rubidium atoms.
 \section*{Funding}
Foundation for Polish Science TEAM Project co-financed by the EU European Regional Development Fund; Foundation for Polish Science Homing Plus project no. 2011-3/14; Polish National Science Centre Project No. 2012/07/B/ST2/00235; National Laboratory FAMO in Toru\'n supported by the subsidy of the Ministry of Science and Higher Education.
%
%
%
%
 \section*{Acknowledgments}
 The authors benefited from numerous discussions  with  colleagues from SYRTE (Paris). We would like to express our particular thanks to S\'ebastien Bize for his very valuable  advice about Hg cooling and trapping. We thank Mateusz Borkowski for careful reading of the manuscript.

 \end{document}